\documentclass[submission,copyright,creativecommons]{eptcs}
\providecommand{\event}{TERMGRAPH 2011} 

\usepackage{graphicx}
\usepackage{multirow}
\usepackage{latexsym}
\usepackage{amsmath,amssymb,amstext}
\usepackage{mathpartir} 
\usepackage{graphicx}
\usepackage{wasysym}
\usepackage{float}
\usepackage{enumerate}
\usepackage[arrow, matrix, curve]{xy}
\usepackage[justification=centering,singlelinecheck=off]{caption}
\usepackage{color}
\usepackage{url}
\usepackage{longtable}
\usepackage{wrapfig}
\usepackage{sidecap}
\usepackage{comment}
\usepackage{MnSymbol}
\usepackage{trfsigns}
\usepackage{fancyvrb}

\usepackage[utf8x]{inputenx}
\usepackage{tipa}

\usepackage{breakurl}             

\newtheorem{theorem}{Theorem} 

\newtheorem{proposition}[theorem]{Proposition}

\newtheorem{definition}[theorem]{Definition}
\newtheorem{remark}[theorem]{Remark}

\newtheorem{notation}[theorem]{Notation}

\newcommand{\lambdaletrec}{\ensuremath{\lambda_\textit{letrec}}}

\newcommand{\sletrec}{\texttt{letrec}}
\newcommand{\letrec}{\texttt{letrec}}
\newcommand{\beIn}{~\texttt{in}~}

\newcommand{\bmetavar}{Y}
\newcommand{\cmetavar}{Z}

\newcommand\listing[1]{\begin{quotation}\noindent\includegraphics{listings/#1}\end{quotation}}
\newcommand\haskell[1]{\ensuremath{\mathit{#1}}}

\newcommand{\slabs}{{\lambda}}

\newcommand{\lbind}[1]{{\slabs{#1}}}
\newcommand{\labs}[2]{\lbind{#1}.\hspace*{0.5pt}{#2}}
\newcommand{\lapp}[2]{{#1}{#2}}

\newcommand{\cxthole}{[\hspace*{0.5pt}]}

\newcommand{\avar}{x}
\newcommand{\bvar}{y}

\newcommand{\afunvar}{f}

\newcommand{\alter}{M}
\newcommand{\blter}{N}
\newcommand{\clter}{P}

\newcommand{\allter}{t}
\newcommand{\bllter}{s}
\newcommand{\cllter}{u}

\newcommand{\slet}{\texttt{let}}

\newcommand{\sletini}[1]{\slet_{#1}\texttt{\_in}}

\newcommand{\letini}[2]{\sletini{#1}{\hspace*{2pt}}{#2}}
\newcommand{\letbeIn}[2]{\slet{\hspace*{2pt}}{#1}{\hspace*{3pt}\texttt{in}\hspace*{5pt}}{#2}}

\newcommand{\atype}{\tau}

\newcommand{\Vars}{\mathit{Var}}
\newcommand{\BaseTypes}{\mathit{BTypes}}
\newcommand{\Types}{\mathit{Types}}

\newcommand{\sred}{\rightarrow}
\newcommand{\red}{\mathrel{\sred}}

\newcommand{\indap}[2]{#1_{#2}}
\newcommand{\sredi}{\indap{\rightarrow}}

\newcommand{\sinvredi}{\indap{\leftarrow}}

\newcommand{\smred}{\twoheadrightarrow}

\newcommand{\smredi}{\indap{\twoheadrightarrow}}

\newcommand{\sminvredi}{\indap{\twoheadleftarrow}}

\newcommand{\thsp}{-1.74ex}
\newcommand{\threeheadrightarrow}{{\twoheadrightarrow\hspace*\thsp\twoheadrightarrow}}
\newcommand{\threeheadleftarrow}{{\twoheadleftarrow\hspace*\thsp\twoheadleftarrow}}

\newcommand{\sinfred}{\threeheadrightarrow}

\newcommand{\sinfredi}{\indap{\threeheadrightarrow}}

\newcommand{\sinvinfredi}{\indap{\threeheadleftarrow}}

\newcommand{\sbetared}{\sredi{\beta}}

\newcommand{\genbetaredsubscript}{g\hspace*{0pt}\beta}
\newcommand{\sgenbetared}{\sredi{\genbetaredsubscript}}

\newcommand{\sinfgenbetared}{\sinfredi{g\hspace*{0pt}\beta}}

\newcommand{\sinvinfgenbetared}{\sinvinfredi{g\hspace*{0pt}\beta}}

\newcommand{\setared}{\sredi{\eta}}
\newcommand{\etared}{\mathrel{\setared}}

\newcommand{\veceta}{\vec{\eta}}
\newcommand{\svecetared}{\sredi{\veceta}}

\newcommand{\vecetazero}{\vec{\eta}_0}
\newcommand{\svecetazerored}{\sredi{\vecetazero}}

\newcommand{\sinvvecetazerored}{\sinvredi{\vecetazero}}

\newcommand{\vecetazeroperm}{\vec{\eta}_{0}^{\text{per}}}
\newcommand{\svecetazeropermred}{\sredi{{\vecetazeroperm}}}

\newcommand{\sinvvecetazeropermred}{\sinvredi{\vecetazeroperm}}

\newcommand{\smvecetazeropermred}{\smredi{{\vecetazeroperm}}}

\newcommand{\sminvvecetazeropermred}{\sminvredi{{\vecetazeroperm}}}

\newcommand{\unfoldredsubscript}{\medtriangledown}
\newcommand{\sunfoldred}{\sredi{\unfoldredsubscript}}
\newcommand{\unfoldred}{\mathrel{\sunfoldred}}

\newcommand{\sinfunfoldred}{\sinfredi{\unfoldredsubscript}}
\newcommand{\infunfoldred}{\mathrel{\sinfunfoldred}}
\newcommand{\sinvinfunfoldred}{\sinvinfredi{\unfoldredsubscript}}
\newcommand{\invinfunfoldred}{\mathrel{\sinvinfunfoldred}}

\newcommand\binds{\mathrel{\scriptsize\fourier\hspace{-0.3em}}}
\newcommand\blackhole{\bullet}

\newcommand{\sopeq}{=^{\infty}_{\unfoldredsubscript,\genbetaredsubscript}}

\newcommand{\sappbisim}{{\sim^{B}}}
\newcommand{\appbisim}{\mathrel{\sappbisim}}

\newcommand{\funap}[2]{{#1}({#2})}
\newcommand{\bfunap}[3]{{#1}({#2,#3})}

\newcommand{\nbd}{\nobreakdash}  
\newcommand{\nbde}{\nobreakdash-\hspace*{0pt}}

\newcommand{\sdefdby}{{:=}}

\newcommand\sep[1]{&#1&} 

\renewcommand\;{\,}

\newcommand{\nats}{\mathbb{N}}

\newcommand\Tau{T}
\newcommand\Rho{R}

\newcommand\T{\ensuremath{\vdash}} 

\newcommand{\saperm}{\pi}
\newcommand{\aperm}{\funap{\saperm}}

\newcommand{\sdiredge}{{\rightarrowtail}}
\newcommand{\diredge}{\mathrel{\sdiredge}}

\newcommand{\srtcdiredge}{{\rightarrowtail^*}}
\newcommand{\rtcdiredge}{\mathrel{\srtcdiredge}\hspace*{-2pt}}

\newcommand{\snotrtcdiredge}{{{\not\rightarrowtail}^*}}
\newcommand{\notrtcdiredge}{\mathrel{\snotrtcdiredge}\hspace*{-2pt}}

\newcommand{\stcdiredge}{{\rightarrowtail^+}}
\newcommand{\tcdiredge}{\mathrel{\stcdiredge}\hspace*{-2pt}}

\newcommand{\snottcdiredge}{{{\not\rightarrowtail}^+}}
\newcommand{\nottcdiredge}{\mathrel{\snottcdiredge}\hspace*{-2pt}}

\newcommand{\sdom}[1]{{{\mit dom}_{#1}}}
\newcommand{\dom}[3]{\bfunap{\sdom{#1}}{#2}{#3}}

\newcommand{\sstrongdom}[1]{{{\mit sdom}_{#1}}}
\newcommand{\strongdom}[3]{\bfunap{\sstrongdom{#1}}{#2}{#3}}

\newcommand{\adigraph}{G}
\newcommand{\verts}{V}

\newcommand{\pair}[2]{\langle {#1},\hspace*{0.5pt} {#2} \rangle}

\newcommand{\aDeriv}{{\cal D}}
\newcommand{\aDerivtilde}{\tilde{\aDeriv}}

\title{Repetitive Reduction Patterns\\ in Lambda Calculus with {\tt letrec}\\
       ({\it Work in Progress})\thanks{Funded by the NWO-project \emph{Realising Optimal Sharing}}}

\author{
  Jan Rochel
    \institute{Utrecht University\\ Utrecht, The Netherlands}
    \institute{Department of Information and Computing Sciences\\
               Information and Software Systems}
    \email{J.Rochel@cs.uu.nl}
\and
  Clemens Grabmayer
    \institute{Utrecht University\\Utrecht, The Netherlands}
    \institute{Department of Philosophy\\
               Theoretical Philosophy}
    \email{Clemens.Grabmayer@phil.uu.nl}
}
\def\titlerunning{Repetitive Reduction Patterns in Lambda Calculus with {\tt letrec}}
\def\authorrunning{Rochel, Grabmayer}

\begin{document}
\maketitle

\begin{abstract}
For the λ-calculus with \texttt{letrec} we develop an optimisation, which
is based on the contraction of a certain class of `future' (also: \emph{virtual})
redexes.

In the implementation of functional programming languages it is common practice
to perform β\nbd-reductions at compile time whenever possible in order to produce
code that requires fewer reductions at run-time. This is, however, in principle
limited to redexes and created redexes that are `visible' (in the sense that
they can be contracted without the need for unsharing), and cannot generally be
extended to redexes that are concealed by sharing constructs such as
\texttt{letrec}. 
In the case of recursion, concealed redexes become visible only after
unwindings during evaluation, and then have to be contracted time and again.

We observe that in some cases such redexes exhibit a certain form of repetitive
behaviour at run time.
We describe an analysis for identifying binders that give rise to such
repetitive reduction patterns, and eliminate them by a sort of predictive
contraction.
Thereby these binders are lifted out of recursive positions or eliminated altogether,
as a result alleviating the amount of β\nbd-reductions required for each recursive iteration. 

Both our analysis and simplification are suitable to be integrated into existing
compilers for functional programming languages as an additional optimisation phase.
With this work we hope to contribute to increasing the efficiency of executing
programs written in such languages.
\end{abstract}

In this extended abstract we report on work in progress carried out within the
framework of the NWO project \emph{Realising Optimal Sharing}. Instead of
discussing optimal reduction in the λ\nbd-calculus, however, here we are concerned
with a static analysis of $\lambdaletrec$-terms which aims at
removing $\beta$\nbd-redexes that are concealed by recursion constructs and
cause cyclic migration of arguments during evaluation. We have to stress that
our research on this particular topic is still in an early phase. 


\section{Introduction}\label{sec:intro}

In this work we study 
terms in $\lambdaletrec$, i.e.\ in $\lambda$\nbde{}calculus with an explicit $\sletrec$\nbde{}construct for recursive definitions,
that exhibit a form of repetitive reduction pattern when evaluated.
We try to identify a class of such terms
for which this behaviour can be avoided by a transformation into a term with,
in some sense, the same semantic denotation.
%
Even though the presented optimisation can be described directly for $\lambdaletrec$\nbd-terms, 
and hence is applicable in all functional languages of which $\lambdaletrec$ is a meaningful abstraction, 
we will use Haskell to denote examples of such terms and their optimised equivalents.
Additionally, we depict terms as λ-graphs with explicit
sharing-nodes (\textit{multiplexers}) as used, for example, in
\cite{AspertiGuerriniOptImpl}.

\begin{figure}[htp]
\begin{center}
\raisebox{-0.5\height}{\includegraphics{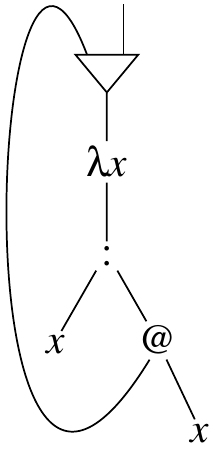}}
\hspace{1mm}$\infunfoldred$\hspace{1mm}
\raisebox{-0.5\height}{\includegraphics{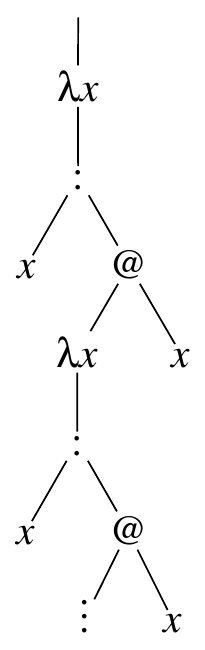}}
\hspace{8mm}$\sinfred_β$\hspace{8mm}
\raisebox{-0.5\height}{\includegraphics{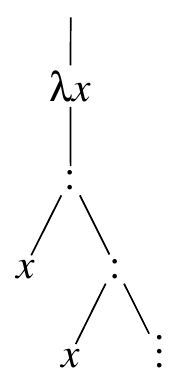}}
\hspace{1mm}$\invinfunfoldred$\hspace{1mm}
\raisebox{-0.5\height}{\includegraphics{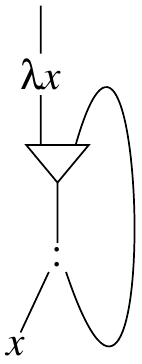}}
\end{center}
\caption{\label{repeat_graphs}Term graph representation of the two \haskell{repeat} implementations and their unfoldings}
\end{figure}

A function well-known to Haskell programmers is the \haskell{repeat} function that
generates an infinite, constant stream of the supplied argument. A definition
is easily found, namely: \listing{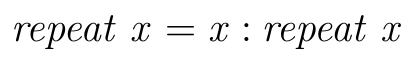}
An experienced Haskell programmer, however, would spot a `space leak',
which refers to an $O(n)$ memory consumption for generating $n$ stream elements
while $O(1)$ is possible, due to lazy evaluation. Therefore in the Haskell
standard libraries that function is defined as: \listing{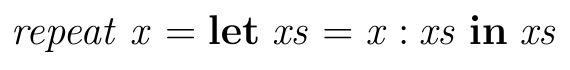}
The exact reasons for this difference in efficiency involve the characteristics
of the deployed Haskell compiler and run-time system. A more direct and
theoretical explanation can be attempted within the 
                                                 formal framework of $\lambdaletrec$:
The improved variant of \haskell{repeat} does not require any β-reductions to
produce further stream elements. That becomes apparent by the λ-graphs and
their infinite unfoldings (Fig. \ref{repeat_graphs}). We use a rewriting
relation arrow indexed by a triangle ($\unfoldred$) to mark unfolding,
regardless of whether sharing is expressed by a multiplexer or a
\letrec-binding.

From a software engineering perspective it is unsatisfactory that the
programmer has to recognise and mitigate such cases. One might even consider
the unoptimised version superior with respect to code
clarity. Therefore we propose an analysis and transformation method to automate
the optimisation, which then can be integrated into the compiler pipeline of
existing functional language implementations.
In the following sections we will work with simple examples to develop this
method and successively generalise it for wider applicability.


%
%
%
%
%

\section{Preliminaries}\label{sec:prelims}

The method we describe applies to the \lambdaletrec-calculus, which is a
higher-order rewrite system. Still, in this work-in-progress report we
primarily intend to motivate our research and outline the approaches developed
so far. Hence, for the moment we give a largely informal description, and
resort to first-order formulations, λ-graphs, or Haskell, whichever seems more suited.

\begin{definition}[First-order representation of $\lambdaletrec$]\normalfont Let $V$ be a set of variables. Then a \lambdaletrec-term is defined as follows:
\begin{equation*}
\begin{array}{lllll}
(\textit{term})        & T    \sep{::=} \lambda V.T         & (\textit{abstraction}) \\
                       &      \sep{ | } T~T                 & (\textit{application}) \\ 
                       &      \sep{ | } V                   & (\textit{variable})    \\
                       &      \sep{ | } \letrec~\mathit{Defs}\beIn T & (\textit{letrec})      \\
(\textit{definitions}) & \mathit{Defs} \sep{::=} v_1=T~\dots~v_n=T   & (\textit{equations})   \\
                       &      \sep{   } v_1,\dots,v_n \in V~\text{all distinct}
\end{array}
\end{equation*}
\end{definition}
%
%

But ultimately only a higher-order formulation can be formally
satisfactory, thus we propose the following representation as a higher-order
rewrite system (HRS) \cite{terese:2003} for \lambdaletrec.

\begin{definition}[Higher-order representation of $\lambdaletrec$]\normalfont
  \label{def:lambdaletrec:terms}
  %
%
  %
  Let $\Vars$ be a set of variables, and $\BaseTypes$ a set of base types 
  that induce the set $\Types$ of simple types.
  The \emph{terms} of $\lambdaletrec$ are simply-typed higher-order terms over the HRS\nbd-signature
  that for all $n\in\nats$, and all types $\atype_0,\atype_1,\ldots,\atype_n\in\Types$ contains a symbol $\sletini{n}$ of type:
  \begin{equation*}
    %
      %
      %
      %
    \sletini{n} \,:\, 
            (\atype_1 \times \ldots \times \atype_n
              \to
             \atype_0 \times \atype_1 \ldots \times \atype_n)    
              \to
            \atype_0
  \end{equation*}
  Product types are only used for better readability.
  We will use the symbols $\allter$, $\bllter$, $\cllter$ for terms in $\lambdaletrec$. 
\end{definition}  
  
Based on this notion of terms, the $\lambdaletrec$\nbd-calculus 
consists of the rewrite relations: $\beta$-reduction, $\eta$\nbd-reduction, and $\letrec$\nbd-unfolding.
Additionally, we use the concept of generalised $\beta$\nbd-reduction \cite{kama:nede:1995}.

For example, $\letrec$\nbd-unfolding on the informal $\lambdaletrec$\nbd-terms according to the grammar above
could be described by the following rewrite rules:
\begin{gather*}
  \letbeIn{\afunvar_1 = \funap{\bllter_1}{\vec{\afunvar}}, \ldots, \afunvar_n = \funap{\bllter_n}{\vec{\afunvar}}}
          {\allter} 
    \;\;\;\sunfoldred\;\;\;
  \allter 
  \qquad \text{(if $\afunvar_1,\ldots,\afunvar_n$ not free in $\allter$)}
  \\[0.5ex]
  \begin{aligned} 
    &
    \letbeIn{\underbrace{\afunvar_1 = \funap{\bllter_1}{\vec{\afunvar}}, \ldots, \afunvar_n = \funap{\bllter_n}{\vec{\afunvar}}}_{\mathit{Defs}}}
            {\funap{\allter}{\vec{\afunvar}}}  
    \;\;\;\sunfoldred\;\;\;
    \allter\bigl(\letbeIn{
                                                               \mathit{Defs}}
                    {\funap{\bllter_1}{\vec{\afunvar}}}, 
    \ldots,  \letbeIn{
                      \mathit{Defs}}
                   {\funap{\bllter_n}{\vec{\afunvar}}}\bigl)
  \end{aligned}
\end{gather*}     
which, if translated into HRS\nbd-rules (using the signature defined in Def.~\ref{def:lambdaletrec:terms}),
can take the following form:  
\begin{gather*}
  \letini{n}{\labs{
                 \afunvar_1\ldots\afunvar_n}{(\bmetavar,\funap{\cmetavar_1}{\vec{\afunvar}},\ldots,\funap{\cmetavar_n}{\vec{\afunvar}})}}
    \;\;\;\sunfoldred\;\;\;
  \bmetavar
  \\[0.5ex]
  \begin{aligned}
    &
    \letini{n}{\labs{
                     \afunvar_1\ldots\afunvar_n}
                    {(\funap{\bmetavar}{\vec{\afunvar}},\funap{\cmetavar_1}{\vec{\afunvar}} ,\ldots, \funap{\cmetavar_n}{\vec{\afunvar}})}
               } 
    \;\;\;\sunfoldred\;\;\;
    \\
    & \hspace*{24ex}
    \bmetavar
    (\letini{n}{\labs{
                   \afunvar_1\ldots\afunvar_n}
               {({\funap{\cmetavar_1}{\vec{\afunvar}}},
                 \funap{\cmetavar_1}{\vec{\afunvar}},\ldots,\funap{\cmetavar_n}{\vec{\afunvar}})}}) \ldots
    \\
    & \hspace*{36ex}
    \phantom{\bmetavar}
    \ldots   
    (\letini{n}{\labs{
                   \afunvar_1\ldots\afunvar_n}
               {({\funap{\cmetavar_n}{\vec{\afunvar}}},\funap{\cmetavar_1}{\vec{\afunvar}},\ldots,\funap{\cmetavar_n}{\vec{\afunvar}})}})          
  \end{aligned}            
\end{gather*}

\begin{notation}[Rewrite relations in $\lambdaletrec$]\normalfont
  On $\lambdaletrec$\nbd-terms we consider the following rewrite relations:  
  $\beta$\nbd-reduction denoted by $\sbetared\;$; generalised $\beta$\nbd-reduction denoted by $\sgenbetared\;$; $\eta$\nbd-reduction denoted by $\setared\;$; $\letrec$\nbd-unfolding denoted by $\sunfoldred$.

  For each of these rewrite relations~$\sred$,  
  the \emph{many-step rewrite relation} with respect to $\sred$ will be written as $\smred$,
  and the (strongly convergent) \emph{infinite rewrite relation} as $\sinfred$.
\end{notation}

With the transformation from Section~\ref{sec:intro}, which is further developed in
the next sections, we aim to convert a given term $\allter$ with repetitive
reduction patterns into a term $\allter'$ that does not require these reductions
to be performed any more, but such that $\allter$ and $\allter'$
are `operationally equivalent', in a sense that guarantees that important properties observable during evaluation are preserved.

One candidate for a precisely defined notion of operational equivalence is
the extension to $\lambdaletrec$\nbd-terms of `applicative bisimulation' on $\lambda$\nbd-terms
due to Abramsky \cite{abra:1990}.
Two $\lambdaletrec$\nbd-terms $\alter$ and $\blter$ are called \emph{applicative bisimilar} (symbolically: $\alter \appbisim \blter$)
if $\alter$ and $\blter$  behave in the same way under all possible series $E_0, E_1, E_2, \ldots$ of `experiments' of the following kind:
on a starting term $\alter_0$ the first experiment $E_0$ consists in finding out whether or not $\alter_0$ reduces to an abstraction (a weak head normal form);
if the outcome $\alter_i$ of the previous experiment $E_i$ is indeed an abstraction $\labs{\avar}{\blter_{i}}$,
then for experiment $E_{i+1}$ an arbitrary term $\clter_{i+1}$ is chosen, and it is determined whether
or not the redex $\lapp{(\labs{\avar}{\blter_{i}})}{\clter_{i+1}}$ reduces to an abstraction.

While applicative bisimulation has been frequently used to justify optimising transformations
for functional programming languages, 
there may be a host of other interesting notions of operational equivalence.
Since, for the moment, we do not want to commit ourselves to a particular notion of operational equivalence, 
we will use a syntactically defined notion of equivalence between terms instead.
In fact, we will define this  syntactic notion as the convertibility relation with respect to 
rewrite relations that we use for motivating and justifying the optimising transformation
in, for example, Fig.~\ref{repeat_graphs} and Fig~\ref{replicate_graphs}.  
There, we use, in addition to infinite convertibility with respect to $\sunfoldred$ and $\sgenbetared$,
a restricted form of `vector $\eta$\nbde{}reduction' 
that is defined by the following rewrite rule:
%
%
\begin{eqnarray*}
  \labs{\avar_1\ldots\avar_n}{\lapp{\alter}{\avar_1\ldots\avar_n}}
      & \red &
  \alter
  \qquad\text{(if $\avar_1, \ldots, \avar_n$ distinct, and not free in $\alter$)}
\end{eqnarray*}
The induced rewrite relation $\svecetared$ extends $\eta$\nbde{}reduction, but can be mimicked with $\eta$\nbd-steps,
and therefore has the same many-step relation.
However, neither for $\eta$\nbde{}reduction nor for vector $\eta$\nbde{}reduction
it holds in generality that the source and the target of a step are applicative bisimilar:
for example, in the $\eta$\nbde{}reduction step $\labs{\avar}{\lapp{\bvar}{\avar}} \etared \bvar$
the source is an abstraction, but the target is not.

Since we want to obtain a syntactically defined notion of operational equivalence that is stronger than applicative bisimilarity,
we define a restriction $\svecetazerored$ of $\svecetared$, and a variant $\svecetazeropermred$ of $\svecetazerored$,
both of which serve our purposes and, importantly, only allow steps between applicative bisimilar terms.
The converse rewrite relation $\sinvvecetazerored$ of $\svecetazerored$ 
performs a copying operation for $\lambda$\nbde{}abstraction prefixes in terms;
and the converse rewrite relation $\sinvvecetazeropermred$ of $\svecetazeropermred$
both copies a $\lambda$\nbde{}abstraction prefix and carries out a permutation in it.

\begin{definition}[Restriction and variant of $\svecetared$]
  \normalfont\label{def:vecetazeropermred}
  The restricted version~$\svecetazerored$ of the rewrite relation $\svecetared$ 
  on $\lambdaletrec$\nbd-terms is defined by the rule:
  \begin{eqnarray*}
    \labs{\avar_1\ldots\avar_n}{\lapp{(\labs{\avar_1\ldots\avar_n}{\alter})}{\avar_1\ldots\avar_n}}
      & \red &
    \labs{\avar_1\ldots\avar_n}{\alter}
      \qquad\text{(if $\avar_1, \ldots, \avar_n$ distinct, and not free in $\alter$)}
  \end{eqnarray*}
  And the extension $\svecetazeropermred$  of $\svecetazerored$ with respect to permuting variable names in abstraction prefixes
  is defined by the rewrite rule: 
  \begin{eqnarray*}
    \labs{\avar_1\ldots\avar_n}{\lapp{(\labs{\avar_{\aperm{1}}\ldots\avar_{\aperm{n}}}{\alter})}{\avar_{\aperm{1}}\ldots\avar_{\aperm{n}}}}
      & \red &
    \labs{\avar_1\ldots\avar_n}{\alter}
    \\
    & & \quad
    \parbox[t]{205pt}{(if $\avar_1, \ldots, \avar_n$ distinct, and not free in $\alter$,
                       \\
                       $\phantom{(\text{if}}$and
                       $\saperm$ is a permutation 
                       on $\{1,\ldots,n\}$)}
  \end{eqnarray*}

\end{definition}
It is easy to verify that left- and right-hand sides of these rules are applicative bisimilar. 

Now we define the syntactic notion of equivalence on which we base our transformation. 

\begin{definition}[Equivalence relation $\sopeq$]%
  \normalfont\label{def:opeq}
  \emph{Infinite convertibility with respect to} $\sunfoldred$ and $\sgenbetared$, extended by finitely many $\svecetazeropermred$\nbde{}reduction steps,
  is the following relation on $\lambdaletrec$\nbd-terms: 
  \begin{equation*}
    \sopeq \;\;\;\sdefdby\;\;\;\;
      (  
         {\sminvvecetazeropermred}
         \; ∪ \;
         {\sinvinfunfoldred} \; ∪ \; {\sinvinfgenbetared}  \; ∪ \; {\sinfgenbetared} \; ∪ \; {\sinfunfoldred} 
         \; ∪ \; 
         {\smvecetazeropermred}  
      )^*
  \end{equation*}
\end{definition}

Since source and targets of each of the rewrite relations $\sunfoldred$, $\sgenbetared$, and $\svecetazeropermred$ 
are applicative bisimilar, and since applicative bisimilarity is a contextual congruence \cite{abra:1990}, 
the following proposition can be proved, which states that the syntactical equivalence 
from Definition~\ref{def:opeq} is at least as strong as, i.e.\ is contained in, applicative bisimulation. 

\begin{proposition}
  For all $\lambdaletrec$\nbd-terms $\alter,\blter$ it holds: {} 
    $\alter \sopeq \blter  \;\Rightarrow\;  \alter \appbisim \blter$.    
\end{proposition}




\section{Further Examples}\label{sec:examples}

The $repeat$ function shows that for some cases it is possible to lift
parameters out of recursive positions and thereby improve run-time efficiency.
That raises the question of when this is possible. What is the pattern that
allows for such an optimisation?

\begin{figure}[htp]
\begin{center}
\raisebox{-0.5\height}{\includegraphics[scale=0.9]{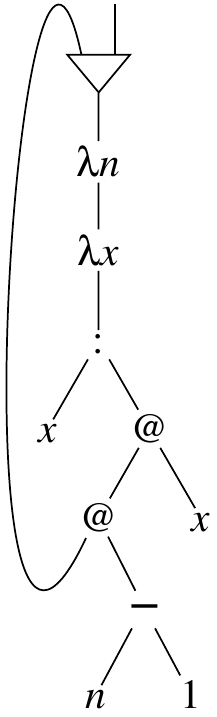}}
$\sinvvecetazerored$
\raisebox{-0.5\height}{\includegraphics[scale=0.9]{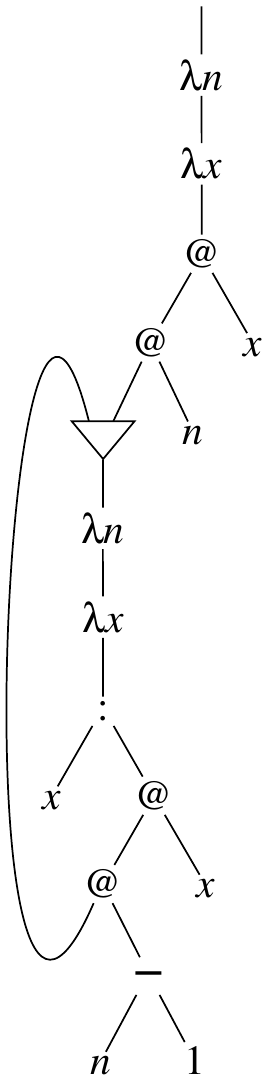}}
$\sinfunfoldred$
\raisebox{-0.5\height}{\includegraphics[scale=0.9]{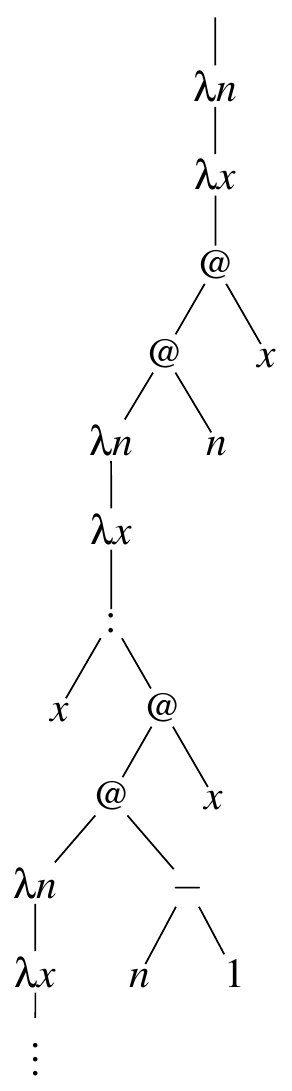}}
$\sinfred_{gβ}$
\raisebox{-0.5\height}{\includegraphics[scale=0.9]{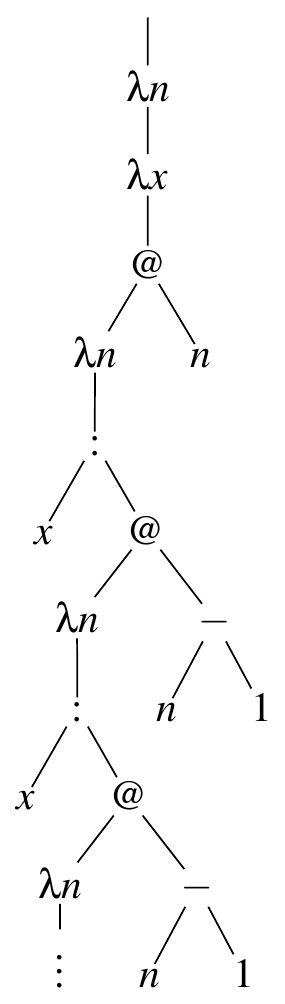}}
$\sinvinfunfoldred$
\raisebox{-0.5\height}{\includegraphics[scale=0.9]{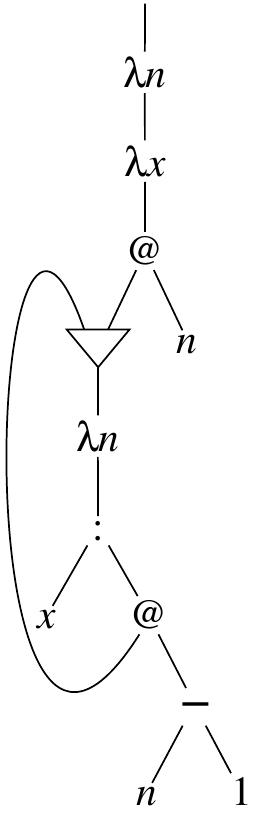}}
\end{center}
\caption{\label{replicate_graphs}Operational equivalence of the two \haskell{replicate} variants depicted in graph notation}
\end{figure}

What strikes the eye are the occurrences of the syntactic element
\haskell{repeat~x} on both the left-hand and the right-hand sides of the
function definition. That suggests a sort of common subexpression elimination
that takes into account both sides of the equation.
This formulation, however, cannot cover the following example, which differs from \haskell{repeat}
essentially only by an additional parameter \haskell{n} on the left-hand side,
and the argument \haskell{n-1} on the right. \listing{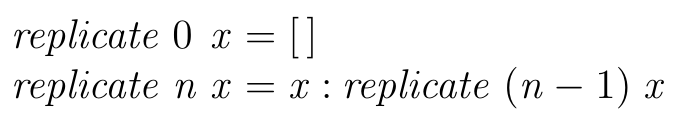}
As it was the case for $\mathit{repeat}$, again it is possible to lift the
parameter $x$ out of the recursion. \listing{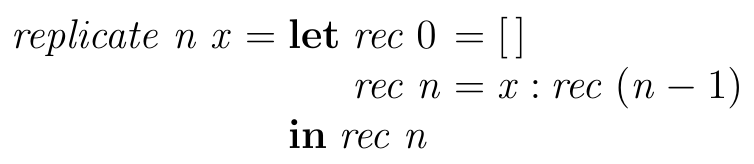}
Again we regard both variants and their infinite unfolding (Fig.
\ref{replicate_graphs}) to understand the transformation. For reasons of
clarity, however, we completely leave out the scrutinisation of
parameter $n$ and the subsequent case discrimination and concentrate on the
recursive pattern.

In comparison with the previous example we observe two differences. First, note
the `header' $λn.λx.~[~]~n~x$ attached on top of the second graph, which we
obtain by two η-expansions. This allows us to produce an optimised term that does
not comprise a duplicated function body. Furthermore, instead of relying
on ordinary β-reduction we have to add {\it generalised beta-reductions} ({\it
gβ}-reduction) \cite{kama:nede:1995} to our arsenal.


The key pattern shared by the two presented examples that permits optimisation
is a parameter $p$ that is being passed through
unchanged in the recursive application. Consequently, once the function is
called from the outside with some argument $a$ in $p$'s position, while
recursively evaluating that call, $p$ can never again be bound to another value
than $a$ or a descendant of $a$. In that sense one might call $p$ a `constant
parameter'. It suggests itself, that components that are `constant' to a
recursive construct can be lifted out of the recursion.

\section{A rewrite rule for simple recursive patterns}
  \label{sec:rules}

The examples in Section~\ref{sec:intro} and Section~\ref{sec:examples} 
suggest that there are many similar situations
in which optimisations of the kind as described can be carried out. 
A first attempt to obtain general formulations of such simplification steps
would be to use schemata, and in effect, rewrite rules on \lambdaletrec\nbd-terms.    

As an example, let us consider the recursive definition of a function $f$ in
which the $(n+1)$th parameter $y$ is passed on to all recursive calls of $f$ as
the $(n+1)$th argument.
In this case the transformation that eliminates the recurrent parameter $y$
can be described by the following first-order rewrite rule on \lambdaletrec
\begin{samepage}
\begin{flushleft}
  \vspace*{-0.5ex}
  \hspace*{6ex}\includegraphics{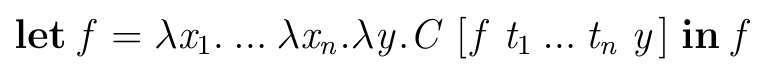}
  \nopagebreak[4]\\[-0.5ex]
  \hspace*{9ex}$\sred$
  \nopagebreak[4]\\[0.5ex]
  \hspace*{6ex}\includegraphics{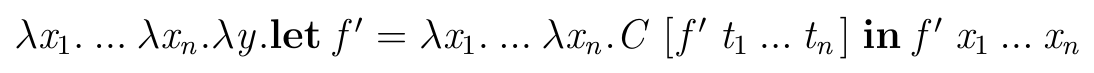}
\end{flushleft}
\end{samepage}
\vspace{-0.5ex}
where $C$ is a \lambdaletrec\nbd-context with possibly more than one occurrence of the context-hole $\cxthole$,
$f$ and $f'$ do not occur in $C$, and $\bvar$ does not get bound during hole-filling.
Here the recurrent parameter $\bvar$ in the recursive definition of $f$ is lifted out
of the recursion, and the number of arguments in the recursive call of the function in
the {\tt let}\nbd-construct has decreased by one after the transformation.
Note, that context $C$ might start off with initial lambdas, and therefore
also covers the case in which $y$ is followed by further parameters.

To cover situations with multiple recursive calls to $f$ with (possibly)
varying arguments, we can generalise the rewrite rule as follows, as long as
the argument in question, $y$, is the same.
\begin{flushleft}
  \vspace*{-0.5ex}
  \hspace*{6ex}\includegraphics{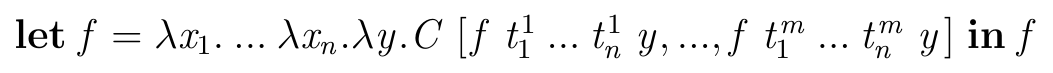}
  \\[-0.5ex]
  \hspace*{9ex}$\sred$
  \\[0.5ex]
  \hspace*{6ex}\includegraphics{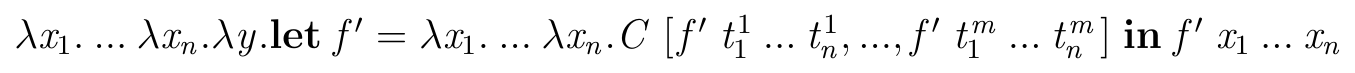}
\end{flushleft}
$C$ is a context with m sort of holes $[]_1$, \ldots, $[]_m$, 
in which holes of each sort may occur more than once,
where $f$ and $f'$ do not occur in $C$, and
the parameter $y$ does not get bound during hole-filling.

In order to enhance its application to a bigger class of \lambdaletrec\nbd-terms,
the second rule can be further generalised
to cover situations in which $f$ is only one amongst many functions defined in
a \letrec\nbd-construct, or there
are also other recursive calls to $f$ that are not of the `good' form. 
%
Namely, if a definition like: \listing{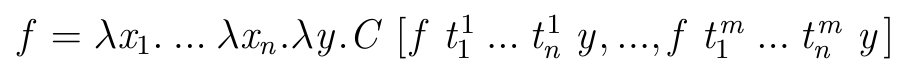} occurs somewhere in a
let-binding, then it can be substituted by: \listing{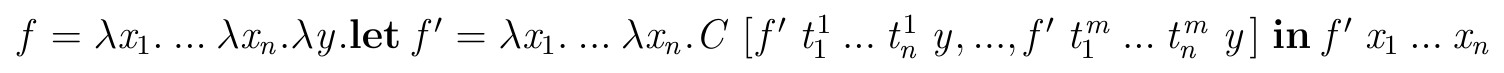}
%
There might be calls of $f$ in $C$ that are of `bad' shape. These remain unchanged. 
On $\lambdaletrec$\nbd-terms, 
this more general transformation can be expressed by the rewrite rule:
\begin{flushleft}
  \vspace*{-0.5ex}
  \hspace*{6ex}\includegraphics{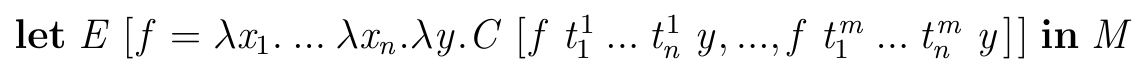}
  \\[-0.5ex]
  \hspace*{9ex}$\sred$
  \\[0.5ex]
  \hspace*{6ex}\includegraphics{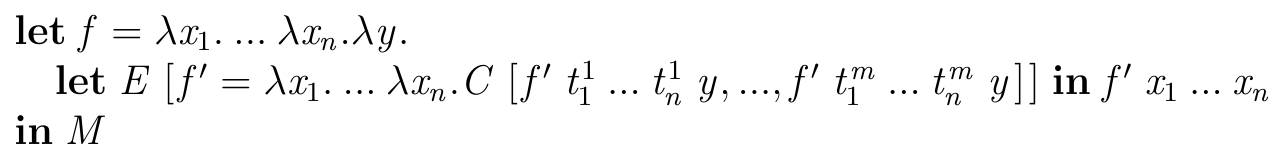}
\end{flushleft}
where $E$ is a context 
of the form $\, D_1, \ldots, D_{i-1}, \cxthole, D_{i}, \ldots, D_l \,$
consisting of definitions $D_{1}$, \ldots, $D_{i-1}$, $D_{i+1}$, \ldots, $D_l$ and a single hole $\cxthole$. 




\subsection{Rewriting the Haskell Prelude}

To demonstrate the above rewriting rule, we apply it to straightforward
implementations of some well-known functions from the Haskell Prelude.

\listing{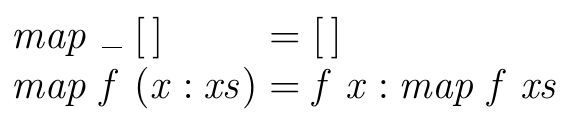}
\listing{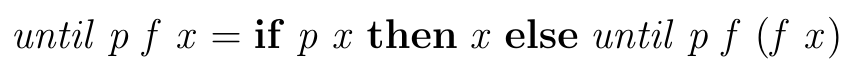}
\listing{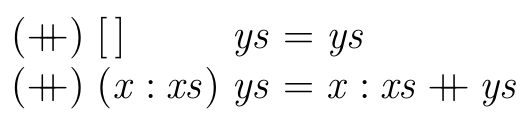}

For the optimised counterparts the amount of β-reduction steps saved per
recursive call amounts to one for \haskell{map} and \haskell{(++)}, and to two
for \haskell{until}.

\listing{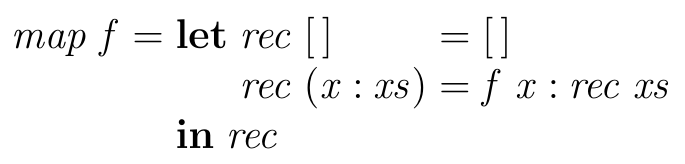}
\listing{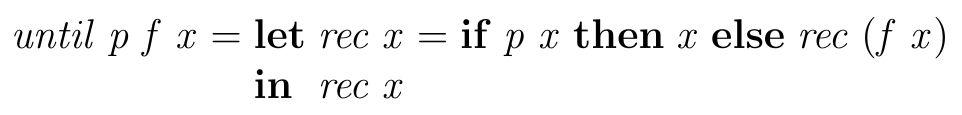}
\listing{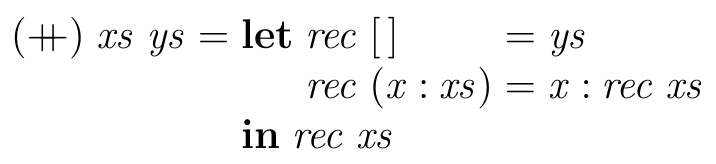}

In practise, however, the amount of β\nbd-reductions is only one of many
factors for the run time that is necessary to evaluate a piece of code.
Therefore it is to be expected that when executed with a system like the
Glasgow Haskell Compiler (GHC) the obtained functions would not necessarily
lead to better performance. Depending on the compiler version and the flags
provided in the invocation of GHC, simple benchmarks yield mixed results, but
never resulting in severe degradation (more than an increase of 5\% in run
time) and with one of the functions (\haskell{until}) consistently reaching a
speed-up of over 200\%.

Let us conclude that before integrating the transformation into a compiler for
practical purposes an analysis on how it interacts with other optimisations
remains yet to be done. 

\subsection{Limitations of the Rewrite Rules}

\begin{figure}[h]
\listing{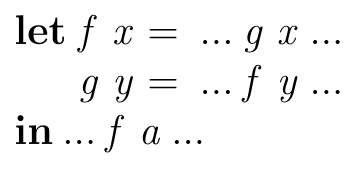}
\caption{\label{mutual}Schematic term involving mutual recursion}
\end{figure}

While the most general rewrite rule above has proven to be applicable in a number of situations
that occur in practice, it also has some severe limitations: 
First and foremost it applies only to patterns with immediate recursion, thus
it fails to capture the repetitive reduction pattern in the evaluation of
the term in Fig. \ref{mutual}.
If \haskell{f} and \haskell{g} are not used at further positions, once in the
recursion initiated by $f~a$ both $x$ and $y$ will never be bound to a
different value than $a$. The property leading to this behaviour is the
relation between the parameters of $f$ and $g$. In $f$, if $g$ is called, then $x$ is
passed as an argument and thereby bound to $y$. Conversely, $y$ is bound to $x$
in the call of $f$ in $g$. Thus, we observe a relation between parameters that
is cyclic. Also for all of the previous example such a {\it parameter cycle}
exists, however comprising only a single parameter. In the following section we
shall elaborate further on the idea of parameter cycles.

\section{Binding Analysis}\label{sec:binding:analysis}

In this section we shall develop an analysis that statically recognises
repetitive reduction patterns indicated by parameter cycles, and show how this
allows us to eliminate those binders that are part of such a cycle. Parameter
cycles describe the possibility of a parameter being passed on from function to
function unchanged, finally arriving at its original position. To detect such
cycles we need to analyse which subterm might be bound to which variable during
the evaluation of a term.



\subsection{Binding Graph}

To this end we introduce a \textit{binding relation} $\binds\ \subseteq V\times
T$ on variables $V$ and subterms $T$ of a term. It is a conservative approximation
of which bindings might occur during the evaluation of the term and does not
distinguish between different descendants of its components.

Since we need to take a global vantage point, i.e.\ to uniquely identify the
term's syntactic elements we assume globally unique variable naming. To this
end we could also use positional information, but only along with additional
technicalities. Therefore, without loss of generality we henceforth assume that
each abstraction binds a distinct variable (variables in the term are `distinctly bound'), 
and no variable name has both a free and a bound occurrence (the term observes `Barendregt's Variable Convention' \cite[2.1.13, p.26]{bare:1984}). 
That allows us to unmistakably address a specific
binding $λx$ by the variable $x$ it binds.\footnote{Even if that forbids
distinguishing between different occurrences of the term $x$, it turns out not
to be necessary for our needs.}

For example for some context $C$ in the reduction of term $t=C[(λx.e_1)~e_2]$
the β-redex $(λx.e_1)~e_2$ might be contracted and by consequence a
descendant of $e_2$ be bound to an instance of $x$. Therefore $t$ implies $x
\binds e_2$. This principle carries over to gβ-reduction, and more importantly,
to sharing.

In order to obtain the binding relation for a term $t$ one has to identify all
terms in argument position that might ever match up with $λx$ in its reduction.
(More precisely: terms whose {\it descendants} might ever match up with {\it
descendants} of $λx$.) This could be naively accomplished by a search that
starts at each abstraction $λx$ and from there travels upwards the spine
segment of $λx$. When an applicator with $a$ as an argument is encountered that
matches $λx$ according to the semantics of gβ-reduction $x\binds a$ is noted
down. When a multiplexer is encountered the search is pursued for each of the
incoming edges.

The use of higher-order functions makes it impossible to enumerate the binding
relation completely. 
Thus, if during the search
the end of the spine is reached one cannot make a safe assumptions on
`future' arguments for that branch. Therefore we employ an additional
`artificial' {\it blackhole} node $\blackhole$ that represents an unknown term
on the right hand side of the binding relation. According to this, the
occurrence of $e_1~(λx.e_2)$ implies $x\binds\blackhole$. Note, that the
blackhole node is not needed to identify parameter cycles, but is however
required for the domination property introduced later on.

Let us revisit the examples presented so far and enlist their binding
relation. In the λ-graph of \haskell{repeat} (Fig. \ref{repeat_graphs}) the
search starting from the only abstraction $λx$ branches at the multiplexer
above. The left branch yields a matching application with $x$ in argument
position, so we obtain $x \binds x$. At right branch the spine immediately
ends yielding $x \binds \blackhole$. The directed graph that is induced by
this relation features a single-node parameter cycle.

\begin{figure}[ht]
\raisebox{-0.5\height}{\includegraphics{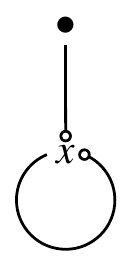}}
\hspace{2cm}
\raisebox{-0.5\height}{\includegraphics{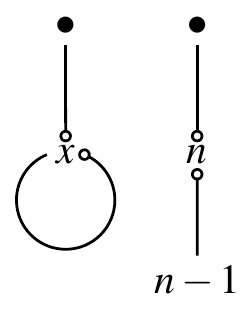}}
\hspace{2cm}
\raisebox{-0.5\height}{\includegraphics{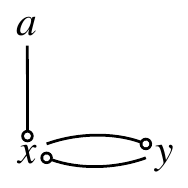}}
\caption{Binding graphs of \haskell{repeat}, \haskell{replicate}, and the term from Fig. \ref{mutual}}
\end{figure}

The binding-graph of \haskell{replicate} is very similar featuring the same
kind of parameter cycle, only does it involve an additional parameter. Also we
have to apply the idea of gβ-reduction when searching upwards from $λx$. First
we encounter another abstraction $λn$, which according to the notion of
gβ-reduction requires us to leave out the next application node. In the left
branch of the multiplexer this is the one with $n-1$ in argument position.
Therefore we end up with $x$ as the argument of the matching application node,
by which we obtain $x \binds x$. Parameter cycles of greater length would occur
for scenarios involving mutual recursion such as the term in Fig. \ref{mutual}.


\subsection{Inference Rules}

To properly define the binding relation we formulate it using inference rules.
Since it follows the structure of typing rules, we first give rules for a
simply-typed $\lambdaletrec$-calculus (Fig. \ref{typing_rules}) and then
decorate these typing rules to also yield a term's binding graph, by which we
hope to provide an easier access for those who are already familiar with typing
rules for the λ-calculus. Type variables are denoted by Greek~letters.

\begin{figure}[htp]
\begin{mathpar}
\inferrule[Var]{x:\tau \in Γ}{Γ ⇒ x:\tau}

\inferrule[Abs]{Γ∪\{x:\tau\} ⇒ e:\sigma}{Γ ⇒ λx.e:\tau→\sigma}

\inferrule[Letrec]{\forall i \in \{0,\dots,n\}: Γ ∪ \{f_1:\tau_1,\dots,f_n:\tau_n\} ⇒ e_i:\tau_i}
                  {Γ ⇒ \letrec ~ f_1 = e_1 ~ \dots ~ f_n = e_n ~ \beIn ~ e_0~:~\tau_0}

\inferrule[App]{Γ ⇒ e_1:\tau→\sigma \\ Γ ⇒ e_2:\rho \\ \tau=\rho}
               {Γ ⇒ e_1~e_2:\sigma}
\end{mathpar}
\caption{\label{typing_rules}Typing rules for the simply-typed $\lambdaletrec$-calculus}
\end{figure}

In order to infer the binding relation (Fig. \ref{annotated}), at any application
$e_1~e_2$ one has to determine to which variable $e_2$ might be bound to in
$e_1$. That we accomplish by annotating the types of terms with the names of
the variables that are bound by their abstractions. The annotations are in
superscript position, so for some annotated types $x:\tau$, $y:\sigma$,
$e_3:\rho$, the annotated type of $e_1 = λx. λy. e_3$ is $e_1 :
(\tau→(\sigma→\rho)^y)^x$. We use $\epsilon$ as an annotation to indicate that
we cannot determine the associated variable (due to the use of higher-order
functions).

The binding relation is denoted on the left-hand side of the turnstile symbol.
Both upper-case and lower-case Greek letters denote annotated type variables,
but the latter do not include the outermost annotation (which are then added
explicitly).

\begin{figure}[h]
\begin{mathpar}
\inferrule[Var]{x:\Tau \in Γ}{\emptyset \T Γ ⇒ x:\Tau}

\inferrule[Abs]{B \T Γ ∪ \{x:\tau^\epsilon\} ⇒ e:\Sigma}{B \T Γ ⇒ λx.e:(\tau^\epsilon→\Sigma)^x}

\inferrule[Letrec]{\forall i \in \{0,\dots,n\}: B_i \T Γ∪\{f_1:\Tau_1, \dots, f_n:\Tau_n\} ⇒ e_i:\Tau_i}
                  {B_0 ∪ \dots ∪ B_n \T Γ ⇒ \letrec ~ f_1 = e_1 ~ \dots ~ f_n = e_n ~ \beIn ~ e_0~:~\Tau_0}

\inferrule[App]{B_1 \T Γ ⇒ e_1:(\Tau→\Sigma)^x \\ B_2 \T Γ ⇒ e_2:\Rho \\ bare(\Tau)=bare(\Rho)}
                {B_1 ∪ B_2 ∪ bind(x,e_2) ∪ blackholes(\Rho) \T Γ ⇒ e_1~e_2:\Sigma}

blackholes(\Tau) =
\begin{cases}
blackholes(\Rho) ∪ \{x\binds\blackhole\} & \text{if $\Tau = (\Sigma→\Rho)^x$} \\
blackholes(\Rho)                         & \text{if $\Tau = (\Sigma→\Rho)^\epsilon$} \\
\emptyset                                & \text{otherwise}
\end{cases}

bind(x,e_2) =
\begin{cases}
\emptyset & \text{if $x = \epsilon$}\\
\{x\binds e_2\} & \text{otherwise}
\end{cases}

\end{mathpar}
\caption{\label{annotated}Inductive definition of the binding relation based on annotated types}
\end{figure}

In the {\sc Abs} rule the abstraction puts a new variable $x$, which is
annotated by an $\epsilon$ because we do not make assumptions on the variables
exposed by higher-order functions. The type of $λx.e$ is annotated by $x$,
which is exposed.
To compare types {\sc App} employs a function $bare: A → T$, which maps
annotated types $A$ to bare types $T$ by removing all annotations. Two further
functions, $bind$ and $blackholes$, are used to enumerate the elements to be
added to the binding relation. In case the argument $e_2$ is a higher-order
function the expressions cannot be determined that are bound to the variables
it exposes, therefore $blackholes$ binds a blackhole to each of them.

\begin{proposition}
  Every derivation $\aDeriv$ in the type system in Fig.~\ref{typing_rules}
  with conclusion $\emptyset \Rightarrow \alter : \tau $,
  which justifies the assignment of type $\tau$ to the $\lambdaletrec$\nbde{}term $\alter$,
  can be decorated effectively with as result
  a derivation $\aDerivtilde$ in the proof system in Fig.~\ref{annotated}
  with conclusion $ B \T \emptyset \Rightarrow \alter : \tilde{\tau} $,
  by which the binding graph $B$ for $\alter$ is obtained and justified.  
\end{proposition}

The proof of this proposition consists in describing an effective algorithm
that, given a derivation $\aDeriv$ in the type system in Fig.~\ref{typing_rules}, proceeds as follows:
First it constructs, in a step by step manner, variable decorations for the types in $\aDeriv$ 
(by concentrating on `spine loops' of $\alter$ that correspond to cyclic threads in $\aDeriv$,
and starting with the decorations at formulas on such threads where the types have minimal length)
in such a way that the rule instances are correct for the system in Fig.~\ref{annotated}
when the leading binding graph annotations of the form $B \T$ are neglected.
Second, after the first step is concluded, the algorithm constructs 
the binding graph annotations according to the rules in Fig.~\ref{annotated}
in a top-down manner.

\section{Transformation}
Using the binding graph we will now develop a more general version of the
optimisation previously formulated as rewriting rules restricted to directly
recursive functions.

An edge $x \binds e$ in the binding graph of a term $t$ indicates a (possibly
infinite) class of gβ-redexes on the infinite unfolding $T$ of $t$. While all
these redexes could be at once contracted on $T$ this does not automatically
carry over to its finite representation $t$.\footnote{In fact, the reduct of an
unfolded $\lambdaletrec$-term does not have to be expressible as a finite term
in $\lambdaletrec$ in general.} Let us for the present restrict our attention
to cases where $x$ has no further incoming edges.

\begin{proposition}\normalfont
Let $t$ be a $\lambdaletrec$-term with infinite unfolding $T$ and a binding
graph featuring a node $x$ with a {\it single incoming} edge $x \binds e$.
If we label the transition that contracts all gβ-redexes with involving a
$λx$-abstraction $red$, and the substitution of all occurrences of $t$ by $e$
and the subsequent elimination of all vacuous $λx$-abstractions $trans$, then
the following diagram commutes.
\[\begin{array}{rcl}
t      & \sinfunfoldred & T \\
trans↓ &                & ↓red \\
t'     & \sinfunfoldred & T'
\end{array}\]
\end{proposition}
This follows from the following considerations. Since we assume unique variable
naming, the infinitely unfolding $t$ without any renaming is semantics
preserving. \cite{endr:grab:klop:oost:2010}\footnote{Even if this has only been
proven for μ-unfolding but it is assumed that it also holds for
$\lambdaletrec$. The unfolding does {\it not} preserve unique naming.} Every
occurrence of $λv$ in $t$ gives rise to one ore more gβ-redexes in $T$ each
having $λv$ with $p$ as an argument. This follows from $x \binds e$ being the
sole incoming edge of $x$. By both paths from $t$ to $T'$ one finds $T'$ to
have the same shape: There are no occurrences of neither $λx$ nor $x$.
Evidently this not a very strong argument and we hope to improve on it by means
of higher-order reasoning.

The restriction above to only consider nodes with a single predecessor prevents
us from dealing with any of the examples shown so far since they involve cyclic
binding graphs. Fortunately, it can be relaxed to a much less restrictive
property.


\begin{definition}[Domination, and strong domination]\normalfont\label{def:domination}
  Let a $ \adigraph = \pair{\verts}{\sdiredge} $ be a directed graph,
  and $u$ and $v$ be vertices of $\adigraph$.
  We say that $v$ \emph{dominates} $w$ ($v$ is a \emph{dominator} for $w$, symbolically: $\dom{\adigraph}{v}{w}$) 
  if either $ v = w$ or
  $v \neq w$ and for every path $\pi$ in $\adigraph$ that leads to $w$ but does not contain $v$
  it holds that the start vertex of $\pi$ is reachable from $v$;
  more formally, if:%
    \footnote{The condition \ref{eq1:def:domination} could be simplified by taking it to be just the
      subformula starting with the universal quantification
      (that subformula is true if $v=w$); the longer condition is used here to increase readability.}
  \begin{equation}\label{eq1:def:domination}
    v = w 
      \;\;\vee\;\;
    \bigl(\,  
    v \neq w \;\;\land\;\;
    \forall u_0, \ldots, u_n\in\verts\setminus\{v\}
      \bigl[\, u_0 \diredge u_1 \diredge \ldots \diredge u_n = w
                 \;\;\Longrightarrow\;\;
               v \rtcdiredge u_0 \,\bigr]
    \bigr)           
  \end{equation}
  %
  Note that $v \rtcdiredge w$ holds if $v$ dominates $w$. 
  
  And we say that $v$ \emph{strongly dominates} $w$ 
  ($v$ is a \emph{strong dominator} for $w$, symbolically: $\strongdom{\adigraph}{v}{w}$)  
  if $v \neq w$ and for every path $\pi$ in $\adigraph$ that leads to $w$ but does not contain $v$
  it holds that the start vertex $u_0$ of $\pi$ is reachable from $v$, but does not reside on a common cycle with $v$,
  more formally, if:
  \begin{equation}\label{eq2:def:domination}
    v \neq w \;\;\land\;\;
    \forall u_0, \ldots, u_n\in\verts\setminus\{v\}
      \bigl[\, u_0 \diredge u_1 \diredge \ldots \diredge u_n = w
                 \;\;\Longrightarrow\;\;
               v \rtcdiredge u_0 
               \;\;\;\land\;\;\;
               u_0 \notrtcdiredge v
               \,\bigr]
  \end{equation}
\end{definition}

\begin{remark}\normalfont
  The standard definition of a `$v$ dominates $w$' for control-flow graphs (see e.g.~\cite{hech:ullm:1974}) 
  requires that each path from the start node to $w$ has to pass through $v$.
  The definition above is a generalisation to directed graphs that does not depend
  on the existence of a designated start node. 
  Our definition of `$v$~strongly dominates $w$' excludes self-domination (i.e.\ makes the relation irreflexive),
  and adds the restriction 
  that for all paths from $v$ to $w$ that do not repeatedly pass through $v$
  it holds no vertex except the starting vertex $v$ resides on a common cycle with~$v$.
\end{remark}

The following proposition suggests an alternative, co-recursive definition of strong domination
between vertices, 
which proceeds stepwisely by examining predecessors of the strongly dominated vertex.

\begin{proposition}\normalfont
  Let $ \adigraph = \pair{\verts}{\sdiredge} $ be a directed graph.
  Then for all $v,w\in\verts$ it holds:
  \begin{equation*}
    \strongdom{\adigraph}{v}{w} 
       \;\;\Longleftrightarrow\;\;
    v \neq w \;\;\land\;\; {v}\tcdiredge{w} \;\;\land\;\; {w}\nottcdiredge{v} 
             \;\;\land\;\;  \forall u\in\verts \bigl(\, u \diredge w \;\;\land\;\; u\neq v \Rightarrow \strongdom{\adigraph}{v}{u} \,\bigr)
  \end{equation*}
\end{proposition}

\vspace{1ex}
For the definition of the optimising transformation, 
we choose a formulation different from the one using rewrite rules described in Section~\ref{sec:rules},
one that is particularly easy to express.
Such as β-reduction can be decomposed into substitution of individual
occurrences of variables (local β-reduction) and the elimination of vacuous
bindings (AT-removal) as described in \cite{brui:1987}, for the transformation
we have the possibility of expressing the transformation with an arbitrary
level of granularity. The following rule encompasses the substitution of all
occurrences of one dominated variable. It could have been formulated more
fine-grained by substituting only individual occurrences, or less fine-grained
by including the elimination of the bindings that have become vacuous.

\begin{proposition}[Main proposition]
In a $\lambdaletrec$-term $t$ occurrences of a variable that is dominated by an
expression $d$ in $t$'s binding graph can be substituted by $d$.
\[\inferrule{B \T \emptyset ⇒ t : \Sigma \and B ⇒ dom_B(d,x)}{t \sopeq t\langle x := d\rangle}\]
\end{proposition}

%
%
%
%
%
%
%
%
%
%
%
%
%
%

\section{Advanced Examples}

\begin{figure}[ht]
\begin{center}
\raisebox{-0.5\height}{\includegraphics{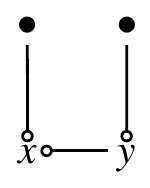}}
\hspace{1cm}
\raisebox{-0.5\height}{\includegraphics{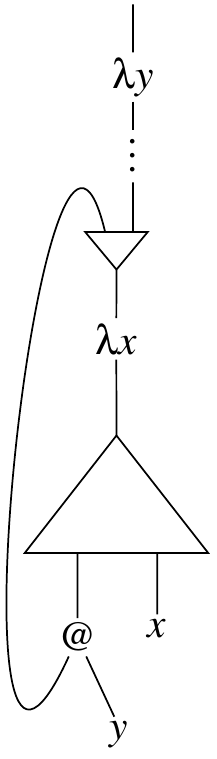}}
\hspace{5mm}$\unfoldred$\hspace{5mm}
\raisebox{-0.5\height}{\includegraphics{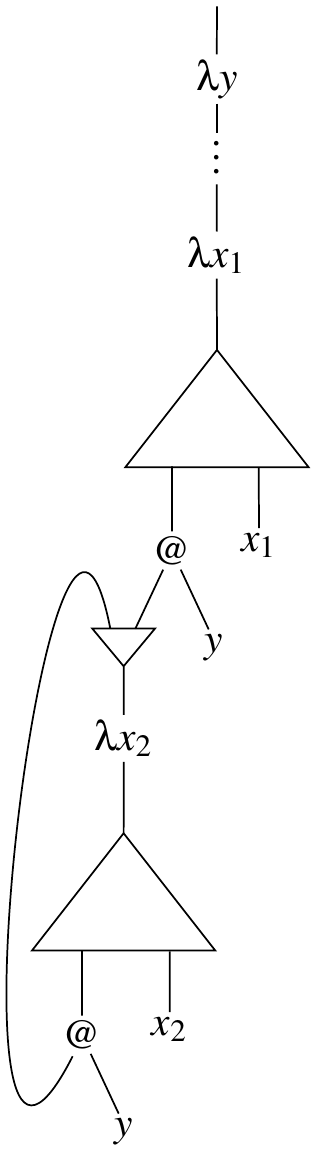}}
\hspace{1cm}
\raisebox{-0.5\height}{\includegraphics{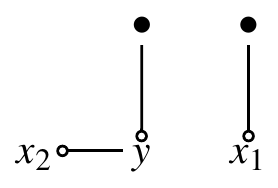}}
\caption{\label{fx-fy}A term, which after one unfolding exhibits a dominated parameter}
\end{center}
\end{figure}

There are interesting examples to which the presented transformation cannot be
directly applied or does not lead to satisfactory results. Only in combination
with a number well-directed unfoldings the desired effect can be obtained. Let
us consider two schematic examples: \listing{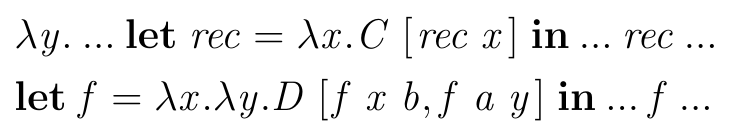} The first one exhibits
dominated parameters only after a single let-unfolding (Fig. \ref{fx-fy}). Also
it features repetitive reduction pattern, without having a parameter cycle.
This because the argument in question is bound outside of the recursion. Still,
after the unfolding the presented optimisation does cure the term.

\begin{figure}[ht]
\raisebox{-0.5\height}{\includegraphics{figs/fx-fy.pdf}}
$\sopeq$
\raisebox{-0.5\height}{\includegraphics{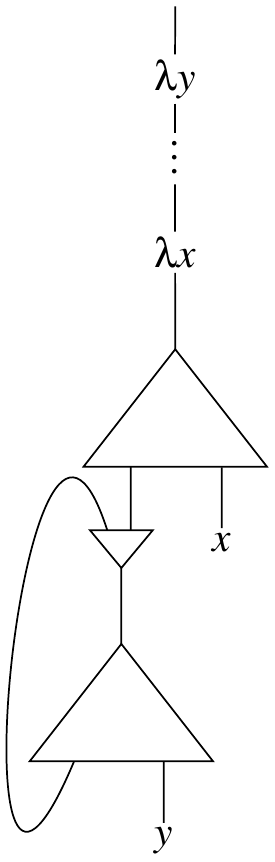}}
\hfill
\raisebox{-0.5\height}{\includegraphics{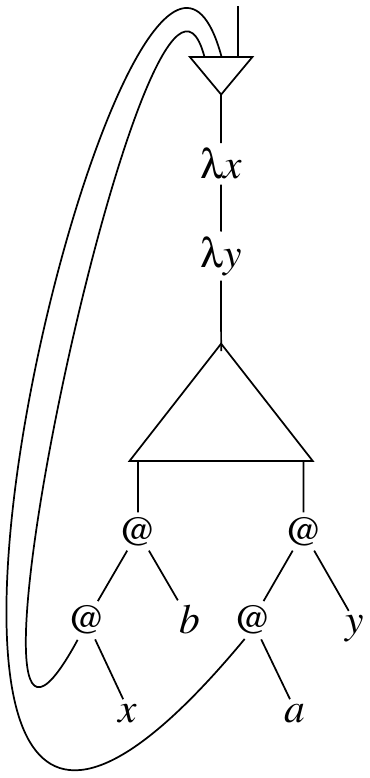}}
$\sopeq$
\raisebox{-0.5\height}{\includegraphics{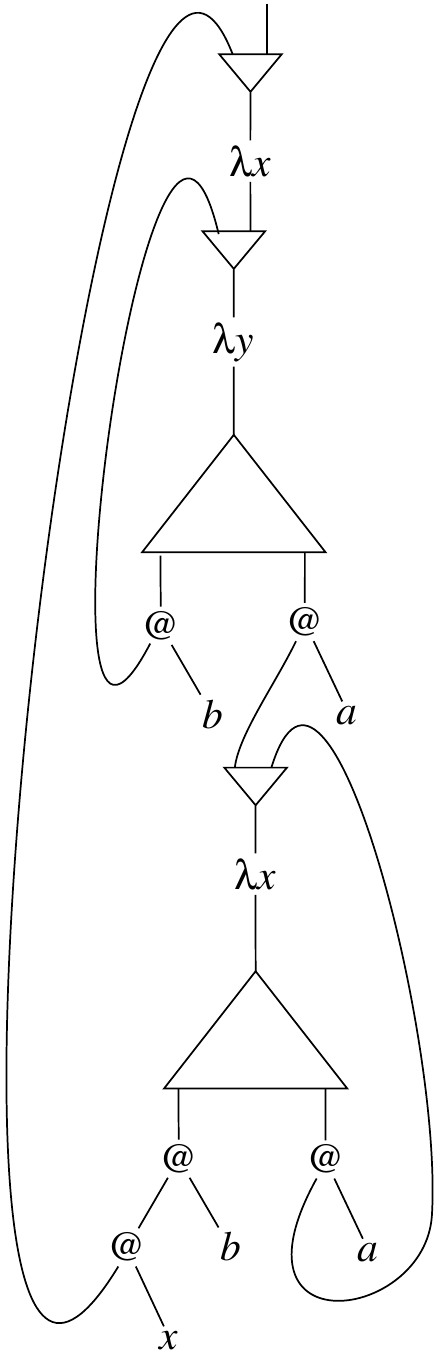}}
\caption{\label{intricate}Optimisation of two terms that are not covered by the presented rewriting rules}
\end{figure}

The second of the two term is particularly intricate since it can be unfolded
and then transformed in many different ways with considerable differences in
the amount of concealed gβ-redexes. Most solutions involve more than one
recursive call with more than two arguments. The ideal transformation with respect to
the number of concealed redexes of both terms is depicted in Fig.
\ref{intricate}.

\section{Status quo}

Presently this work consists mostly of an extended problem description, which
is to a some extent rather informal. Ideas for resolution have been presented,
but yet lack both precision and genericity. Therefore, outstanding issues to
continue the investigation suggest themselves. Currently we are working on the
following problems:
\begin{itemize}
  \item Properly formalising the concepts introduced in this report. This
  involves putting grammars and rewrite rules into the higher-order setting of HRSs. 
  \item Investigate whether there are other interesting notions 
  of `operational equivalence', apart from applicative bisimulation,
  with respect to which we could show correctness of our optimising transformation.
  (We think of notions that are in line with the semantics of
   functional programming languages, but nevertheless are language independent,
   and also of theoretical interest.)
  \item Expressing the presented transformation as a higher-order rewrite
  system and proving its correctness with respect to that equivalence relation.
\end{itemize}

Once these fundamental issues are resolved, we intend to tackle the following questions: 
\begin{itemize}
  \item We have seen that unfolding a \lambdaletrec-term in order to facilitate
  the optimisation can be effected in different ways, leading to terms of
  different quality. To obtain the most efficient terms one has to provide a
  procedure to select the most suitable unfolding for each situation.
  \item This requires an adequate measure for efficiency.
  \item We would like to learn more about the rewrite properties of the transformation:
    is it possible to find a formulation that guarantees confluence and normalisation?
  \item In some of the easy examples we studied, our transformation seemed to be
    closely connected with the concept of `lambda-dropping' \cite{danv:schu:1997,danv:1999},
    and hence also with its converse, `lambda-lifting' \cite{john:1985,peyt:jone:1987,danv:schu:2002}.
    We want to understand that relationship in detail. 
  \item Once the analysis has been optimised in terms of genericity, i.e.\ that
  it recognises as many cases as possible for which the transformation is
  correct, it would be interesting to assess the frequency in which these cases
  occur in existing systems, such as functional programming libraries or
  intermediate code generated by compilers.
  \item The remaining question is, how the optimisation actually affects the
  run-time efficiency of real-world systems like Haskell programs.
\end{itemize}





\paragraph{Acknowledgment.}
  The incentive to investigate the presented optimisation was provided by
  Doaitse Swierstra. We thank him and Vincent van Oostrom for many insightful
  discussions and hints.

\bibliographystyle{eptcs}
\bibliography{termgraph-2011}

\begin{thebibliography}{10}
\providecommand{\bibitemdeclare}[2]{}
\providecommand{\urlprefix}{Available at }
\providecommand{\url}[1]{\texttt{#1}}
\providecommand{\href}[2]{\texttt{#2}}
\providecommand{\urlalt}[2]{\href{#1}{#2}}
\providecommand{\doi}[1]{doi:\urlalt{http://dx.doi.org/#1}{#1}}
\providecommand{\bibinfo}[2]{#2}

\bibitemdeclare{inproceedings}{abra:1990}
\bibitem{abra:1990}
\bibinfo{author}{Samson Abramsky} (\bibinfo{year}{1990}):
  \emph{\bibinfo{title}{The Lazy Lambda Calculus}}.
\newblock In: {\sl \bibinfo{booktitle}{Research Topics in Functional
  Programming}}. \bibinfo{publisher}{Addison-Wesley}, pp.
  \bibinfo{pages}{65--116}.
\newblock \bibinfo{note}{Updated version (2006) available at
  \url{http://web.comlab.ox.ac.uk/people/Samson.Abramsky/lazy.pdf}}.

\bibitemdeclare{book}{AspertiGuerriniOptImpl}
\bibitem{AspertiGuerriniOptImpl}
\bibinfo{author}{Andrea Asperti} \& \bibinfo{author}{Stefano Guerrini}
  (\bibinfo{year}{1998}): \emph{\bibinfo{title}{The Optimal Implementation of
  Functional Programming Languages}}.
\newblock {\sl \bibinfo{series}{Cambridge Tracts in Theoretical Computer
  Science}}~\bibinfo{volume}{45}, \bibinfo{publisher}{Cambridge University
  Press}.

\bibitemdeclare{book}{bare:1984}
\bibitem{bare:1984}
\bibinfo{author}{H.P. Barendregt} (\bibinfo{year}{1984}):
  \emph{\bibinfo{title}{{The Lambda Calculus: Its Syntax and Semantics}}},
  \bibinfo{edition}{$2$nd} edition.
\newblock {\sl \bibinfo{series}{SLFM}} \bibinfo{volume}{103},
  \bibinfo{publisher}{Elsevier}.

\bibitemdeclare{techreport}{brui:1987}
\bibitem{brui:1987}
\bibinfo{author}{N.G. de~Bruijn} (\bibinfo{year}{1987}):
  \emph{\bibinfo{title}{{Generalizing Automath by Means of a Lambda-Typed
  Lambda Calculus}}}.
\newblock \bibinfo{type}{Technical Report} \bibinfo{number}{AUT092 (AUTOMATH
  archive \url{http://www.win.tue.nl/automath/})},
  \bibinfo{institution}{Technische Universiteit Eindhoven, Eindhoven, the
  Netherlands}.
\newblock
  \urlprefix\url{http://alexandria.tue.nl/repository/freearticles/597608.pdf}.

\bibitemdeclare{incollection}{danv:1999}
\bibitem{danv:1999}
\bibinfo{author}{Olivier Danvy} (\bibinfo{year}{1999}):
  \emph{\bibinfo{title}{An Extensional Characterization of Lambda-Lifting and
  Lambda-Dropping}}.
\newblock In \bibinfo{editor}{Aart Middeldorp} \& \bibinfo{editor}{Taisuke
  Sato}, editors: {\sl \bibinfo{booktitle}{Functional and Logic Programming}}.
  {\sl \bibinfo{series}{LNCS}} \bibinfo{volume}{1722},
  \bibinfo{publisher}{Springer Berlin / Heidelberg}, pp.
  \bibinfo{pages}{241--250}, \doi{10.1007/10705424\_16}.

\bibitemdeclare{incollection}{danv:schu:2002}
\bibitem{danv:schu:2002}
\bibinfo{author}{Olivier Danvy} \& \bibinfo{author}{Ulrik Schultz}
  (\bibinfo{year}{2002}): \emph{\bibinfo{title}{Lambda-Lifting in Quadratic
  Time}}.
\newblock In \bibinfo{editor}{Zhenjiang Hu} \& \bibinfo{editor}{Mario
  Rodr\'{i}guez-Artalejo}, editors: {\sl \bibinfo{booktitle}{Functional and
  Logic Programming}}. {\sl \bibinfo{series}{LNCS}} \bibinfo{volume}{2441},
  \bibinfo{publisher}{Springer Berlin / Heidelberg}, pp.
  \bibinfo{pages}{134--151}, \doi{10.1007/3-540-45788-7\_8}.

\bibitemdeclare{inproceedings}{danv:schu:1997}
\bibitem{danv:schu:1997}
\bibinfo{author}{Olivier Danvy} \& \bibinfo{author}{Ulrik~P. Schultz}
  (\bibinfo{year}{1997}): \emph{\bibinfo{title}{Lambda-dropping: transforming
  recursive equations into programs with block structure}}.
\newblock In: {\sl \bibinfo{booktitle}{Proceedings of the 1997 ACM SIGPLAN
  symposium on Partial evaluation and semantics-based program manipulation}}.
  \bibinfo{series}{PEPM '97}, \bibinfo{publisher}{ACM}, \bibinfo{address}{New
  York, NY, USA}, pp. \bibinfo{pages}{90--106}, \doi{10.1145/258993.259007}.

\bibitemdeclare{unpublished}{endr:grab:klop:oost:2010}
\bibitem{endr:grab:klop:oost:2010}
\bibinfo{author}{J\"{o}rg Endrullis}, \bibinfo{author}{Clemens Grabmayer},
  \bibinfo{author}{Jan~Willem Klop} \& \bibinfo{author}{Vincent van Oostrom}
  (\bibinfo{year}{2010}): \emph{\bibinfo{title}{{On Equal
  $\mu$\protect\nbde{}Terms}}}.
\newblock \bibinfo{note}{Submitted, currently under review, to be published in
  2011}.

\bibitemdeclare{article}{hech:ullm:1974}
\bibitem{hech:ullm:1974}
\bibinfo{author}{M.~S. Hecht} \& \bibinfo{author}{J.~D. Ullman}
  (\bibinfo{year}{1974}): \emph{\bibinfo{title}{Characterizations of Reducible
  Flow Graphs}}.
\newblock {\sl \bibinfo{journal}{JACM}} \bibinfo{volume}{21}, pp.
  \bibinfo{pages}{367--375}, \doi{10.1145/321832.321835}.

\bibitemdeclare{incollection}{john:1985}
\bibitem{john:1985}
\bibinfo{author}{Thomas Johnsson} (\bibinfo{year}{1985}):
  \emph{\bibinfo{title}{Lambda lifting: Transforming programs to recursive
  equations}}.
\newblock In \bibinfo{editor}{Jean-Pierre Jouannaud}, editor: {\sl
  \bibinfo{booktitle}{Functional Programming Languages and Computer
  Architecture}}. {\sl \bibinfo{series}{LNCS}} \bibinfo{volume}{201},
  \bibinfo{publisher}{Springer Berlin / Heidelberg}, pp.
  \bibinfo{pages}{190--203}, \doi{10.1007/3-540-15975-4\_37}.

\bibitemdeclare{article}{kama:nede:1995}
\bibitem{kama:nede:1995}
\bibinfo{author}{Fairouz Kamareddine} \& \bibinfo{author}{Rob Nederpelt}
  (\bibinfo{year}{1995}): \emph{\bibinfo{title}{Refining reduction in the
  lambda calculus}}.
\newblock {\sl \bibinfo{journal}{Journal of Functional Programming}}
  \bibinfo{volume}{5}(\bibinfo{number}{4}), pp. \bibinfo{pages}{637--651},
  \doi{10.1017/S0956796800001507}.

\bibitemdeclare{book}{peyt:jone:1987}
\bibitem{peyt:jone:1987}
\bibinfo{author}{Simon~L. Peyton~Jones} (\bibinfo{year}{1987}):
  \emph{\bibinfo{title}{The Implementation of Functional Programming Languages
  (Prentice-Hall International Series in Computer Science)}}.
\newblock \bibinfo{publisher}{Prentice-Hall, Inc.}, \bibinfo{address}{Upper
  Saddle River, NJ, USA}.

\bibitemdeclare{book}{terese:2003}
\bibitem{terese:2003}
\bibinfo{author}{Terese} (\bibinfo{year}{2003}): \emph{\bibinfo{title}{{Term
  Rewriting Systems}}}.
\newblock {\sl \bibinfo{series}{Cambridge Tracts in Theoretical Computer
  Science}}~\bibinfo{volume}{55}, \bibinfo{publisher}{Cambridge University
  Press}.

\end{thebibliography}
\end{document}